\documentclass[11pt,fleqn]{article}

\usepackage{amsmath,amssymb}
\usepackage{graphicx}
\usepackage[top=0.75in, bottom=0.75in, left=0.75in, right=0.75in, dvips]{geometry}

\newcommand{\dif}[1]{\ensuremath{ \mathrm{d}#1 }}

\newcommand{\e}{\ensuremath{\mathrm{e}}}

\newcommand{\ii}{\ensuremath{\mathrm{i}}}

\def\tref#1#2#3{{#1}~(#2)~#3}

\newcommand{\ie}{\textit{i.e.}}
\newcommand{\nf}{ \ensuremath{n_\text{f}} }
\newcommand{\alpi}{ \frac{\alpha}{\mathrm{\pi}} }
\newcommand{\dDx}{ \mathrm{d}^{D}\! x }
\newcommand{\normal}[1]{: \! #1 \! :}
\newcommand{\ket}[1]{\ensuremath{| #1\rangle} }
\newcommand{\bracket}[3]{ \ensuremath{\langle #1| #2 |#3\rangle} }
\newcommand{\mc}[1]{ \ensuremath{\mathcal{#1}} }
\newcommand{\tprod}{ \ensuremath{\mathcal{T}} }
\newcommand{\Ok}[1]{ \ensuremath{\mathcal{O}(k_{1,2}^{#1})} }
\newcommand{\polarization}%
  { \left[(k_1 \cdot k_2)(\epsilon_1 \cdot\epsilon_2)
         -(k_1 \cdot\epsilon_2)(k_2 \cdot\epsilon_1)\right] }
\newcommand{\mathfig}[2]%
{
  \raisebox{ -0.5\totalheight}%
  {\includegraphics[scale=#2]{#1}}
}

\begin{document}
\title{{\sf Quark Effects in the Gluon Condensate Contribution to the
    Scalar Glueball Correlation Function}}
\author{
  D.~Harnett\thanks{email: derek.harnett@ucfv.ca}\\
\textsl{Department of Physics}\\
\textsl{University College of the Fraser Valley}\\
\textsl{Chilliwack, British Columbia, V2P 6T4, Canada}\\[1em]
T.G.~Steele\thanks{email: Tom.Steele@usask.ca}\\
  \textsl{Department of Physics and Engineering Physics}\\
  \textsl{University of Saskatchewan}\\
  \textsl{Saskatoon, Saskatchewan, S7N 5E2, Canada}\\
 }

\maketitle
\begin{abstract}
One-loop quark contributions to the dimension-four gluon condensate term
in the operator product expansion (OPE) of the scalar
glueball correlation function are calculated in the $\overline{\text{MS}}$ scheme 
in the chiral limit of $\nf$ quark flavours. 
The presence of quark effects is shown not to alter the cancellation of infrared
(IR) singularities in the gluon condensate OPE coefficients.
The dimension-four gluonic condensate term represents the leading power corrections
to the scalar glueball correlator and, therein, the one-loop logarithmic contributions
provide the most important condensate contribution to
those QCD sum-rules independent of the low-energy theorem (the subtracted sum-rules).
\end{abstract}

The QCD correlation function of scalar gluonic currents
\begin{gather}
  \Pi\left(q^2\right)=
  \ii\int\!\dDx\;\e^{\ii q\cdot x} \bracket{0}{\tprod J(x)J(0)}{0},
\label{corr_fn}\\
  J(x)=\alpha G^a_{\mu\nu}(x)G^{a\mu\nu}(x)\equiv\alpha G^2(x),
\label{current}
\end{gather}
is used to study the properties of scalar gluonium via QCD sum-rule
techniques~\cite{glue_sr1,glue_sr2}.
The current $J(x)$ is the lowest-order version of the operator
$\beta(\alpha) G^2(x)$, which is renormalization-group invariant for chiral
quarks~\cite{RG,op_mix}. As first noted in~\cite{bagan_ir,bagan_sr},
the one-loop gluon condensate contribution to~(\ref{corr_fn})
\begin{equation}\label{d4cond}
  \left[a_0 +b_0\alpi+b_1\alpi
  \log\left(\frac{-q^2}{\nu^2}\right)\right]
  \left\langle\alpha G^2\right\rangle,
\end{equation}
where $\{a_0,\, b_0,\, b_1\}$ are numerical coefficients
(see~(\ref{finalresultsa0})--(\ref{finalresultsb1}) below),
represents the leading condensate contribution to those sum-rules independent of the
low-energy theorem~\cite{let}, and so provide important non-perturbative effects
within sum-rule analyses of scalar glueballs.

The one-loop coefficients $b_0$ and $b_1$ were evaluated in~\cite{bagan_ir}
in the absence of quark effects (\ie\ the $\nf=0$
limit). As these $\nf=0$ gluon condensate effects have been used in a number of sum-rule
analyses where the effects of three quark flavours have been
included in the perturbative part~\cite{glue_sr2},
it is  necessary to extend the results
 of~\cite{bagan_ir} to enable self-consistent sum-rule analyses in the presence of $\nf$ chiral quarks.

Since the gluonic current~(\ref{current}) is gauge invariant,
the operator product expansion (OPE) of the corresponding
two-point operator contains only those local operators which are
gauge invariant, equations of motion, or BRS variations~\cite{ope};
hence, for massless quarks, the relevant OPE is given by
\begin{multline}
  \ii\int\!\dDx\;\e^{\ii q\cdot x}\tprod G^2(x)G^2(0)\\
  = \mathcal{I}(q^2) + \mathcal{C}(q^2) \normal{G^2(0)}
  + \mathcal{D}(q^2)q^{\sigma}q^{\lambda}\left[\normal{G^{a}_{\ \mu\sigma}(0)
    G^{a\mu}_{\ \ \lambda}(0)}-\frac{1}{D}g_{\sigma\lambda}\normal{G^2(0)} \right]\\
  + \text{equation of motion \& BRS invariant operators}
  + \text{higher dimensional operators}
\label{ope}
\end{multline}
where $\mc{I}(q^2),\,\mc{C}(q^2),\,\mc{D}(q^2)$ are the Wilson (OPE)
coefficients and where the two colons indicate normal ordering.  For notational convenience,
the normal ordering symbol and the spacetime argument will subsequently be omitted from
the right-hand side of~(\ref{ope}).

The scalar gluonic correlator~(\ref{corr_fn}) is obtained from~(\ref{ope})
through multiplication by $\alpha^2$ followed by a vacuum expectation value (VEV).
Physical matrix elements of the equation of motion and BRS invariant
operators vanish, as does the VEV of the gluonic operator proportional to $\mc{D}(q^2)$.
Therefore, up to dimension four operators, the
sole contributions to~(\ref{corr_fn}) stem from $\mc{I}(q^2)$
(perturbation theory) and $\mc{C}(q^2)$.  This article is concerned with the latter.

As in~\cite{bagan_ir}, the nonzero momentum insertion technique
(NZI method)~\cite{chetyrkin,bagan_nzi} is employed to compute  OPE coefficients.
This method sandwiches the OPE between two external single gluon states with momenta $k_1\ne k_2$, permitting the simple separation of
(non-physical) operators whose VEV is zero (\ie\ BRS invariants and equations of motion) from the physical operator $G^2$,
thus simplifying the operator-mixing effects originating from the renormalization of composite operators \cite{op_mix}.  The NZI method
also facilitates the analysis of infrared aspects of the OPE, since the nonzero difference between the external gluon momenta
$s=\left(k_1-k_2\right)^2$ provides an infrared regulator.  As shown in \cite{bagan_nzi} for the $\nf=0$ case, all infrared logarithms
(\ie\ $\log(s)$, $\log{\left(k_1^2\right)}$, and $\log{\left(k_2^2\right)}$) cancel in the calculation of the OPE coefficients.
This then allows the use of on-shell external gluon states
\begin{equation}
\ket{n}\equiv\ket{\epsilon_n,\,k_n,\,c_n},~\epsilon_n\cdot k_n=k_n^2=0
\label{states}
\end{equation}
to sandwich the OPE, immediately eliminating all non-physical operators.  The colour index $c_n$ associated with the gluon states
(\ref{states}) is  included for completeness, but only leads to trivial overall colour factors when taking matrix elements of
colour singlet objects such as (\ref{ope}).  As will be discussed below, the IR cancellation in the OPE coefficients is not
altered by the inclusion of quarks, and hence these simplifications can be applied even in the presence of chiral quarks.

Sandwiching~(\ref{ope}) between on-shell gluon states yields
\begin{multline}\label{beginning}
  \ii\int\!\dDx\;\e^{\ii q\cdot x}\bracket{1}{\tprod G^2(x)G^2(0)}{2}
  = \mc{C}(q^2)\bracket{1}{G^2}{2}
  + \mc{D}(q^2)q^{\sigma}q^{\lambda}
    \bracket{1}{G^{a}_{\ \mu\sigma}G^{a\mu}_{\ \ \lambda}
    -\frac{1}{D}g_{\sigma\lambda}G^2}{2}\\
  +\text{contributions from higher dimension ($d>4$) operators},
\end{multline}
where the matrix elements of all non-physical operators have been eliminated
through the use of on-shell external gluon states.
To simulate the effects of a VEV,
the direction averaging operator $\int\!\dif{\hat{q}}$ which, for example,
leads to the identity
\begin{equation}\label{vacuumavg}
  \int\!\dif{\hat{q}}\; q_{\alpha}q_{\beta} f(q^2)=\frac{1}{D}q^2
  g_{\alpha\beta} f(q^2),
\end{equation}
is applied to~(\ref{beginning}).
Consequently, the term proportional to $\mc{D}(q^2)$ is annihilated (as in the VEV).
In addition, it should be noted that $\bracket{1}{G^2}{2}=\Ok{2}$ whereas the
remaining terms on the right-hand side of~(\ref{beginning}) go like $\Ok{4}$.
In this way, contributions relevant to the computation of $\mc{C}(q^2)$ are easily
identified.  Therefore, Eq.~(\ref{beginning}) implies
\begin{equation}\label{intermediate}
  \ii\int\!\dif{\hat{q}}\,\dDx\;\e^{\ii q\cdot x}\bracket{1}{\tprod G^2(x)G^2(0)}{2}
  = \mc{C}(q^2)\bracket{1}{G^2}{2} +\Ok{4}.
\end{equation}
Thus, at leading and next-to-leading order
(respectively denoted by the $(0)$ and $(1)$ superscripts in what follows)
in the bare fields, we have the set of equations
\begin{align}
  \ii\int\!\dif{\hat{q}}\,\dDx\;\e^{\ii q\cdot x}\bracket{1}{\tprod G_B^2(x)G_B^2(0)}{2}^{(0)}
  &=\mc{C}^{(0)}(q^2) \bracket{1}{G_B^2}{2}^{(0)} +\Ok{4},\label{lo}\\
  \ii\int\!\dif{\hat{q}}\,\dDx\;\e^{\ii q\cdot x}\bracket{1}{\tprod G_B^2(x)G_B^2(0)}{2}^{(1)}
  &=\mc{C}^{(0)}(q^2) \bracket{1}{G_B^2}{2}^{(1)}
   +\mc{C}^{(1)}(q^2) \bracket{1}{G_B^2}{2}^{(0)}+\Ok{4},\label{nlo}
\end{align}
where the subscript $B$ denotes a perturbative expansion in terms of bare
(unrenormalized) quantities.

Consider first~(\ref{lo}) which, in terms of Feynman diagrams, is given schematically by
\begin{equation}\label{feynmanlo}
  \ii\int\!\dif{\hat{q}}
  \left\{ \mathfig{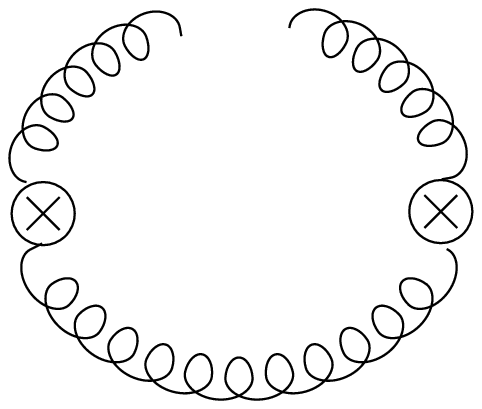}{0.5} +
  {\rm crossed~diagram}
  \right\}
  =\mc{C}^{(0)}(q^2)\left\{\mathfig{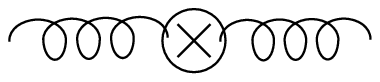}{0.5}\right\}
  +\Ok{4}.
\end{equation}
which implies
\begin{multline}\label{losimplified}
  \frac{64}{D}\delta^{c_1 c_2}\polarization +\Ok{4}\\
  =4\mc{C}^{(0)}(q^2)\delta^{c_1 c_2}\polarization +\Ok{4}.
\end{multline}
Therefore
\begin{equation}\label{c0}
  \mc{C}^{(0)}(q^2)=\frac{16}{D},
\end{equation}
and we note that, at leading order, $\mc{C}(q^2)$ is unaffected by the inclusion of
chiral quarks. The spacetime dimension $D$ must be kept arbitrary until later stages of the calculations.

\begin{figure}[htb]
\centering
\includegraphics[scale=0.6]{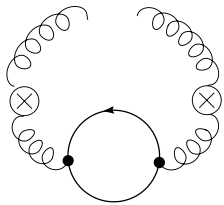}
\caption{Feynman diagram containing quark effects contributing to the amplitude on the left-hand side of
(\protect\ref{nlo}).}
\label{quarkfig}
\end{figure}

Quark contributions to $\mc{C}(q^2)$ do, however, show up at next-to-leading order,
but \emph{only} through that contribution to the left-hand side of~(\ref{nlo}) stemming from the
diagram (and its crossed counterpart) depicted in Figure~\ref{quarkfig}.
This diagram does not generate any IR singularities in the OPE analysis 
since the external gluon momenta regulate the IR behaviour associated with the quark loop.  
Hence, the conclusion that all IR singularities cancel in the
calculation of the gluon condensate OPE coefficient~\cite{bagan_ir} 
is upheld in the presence of chiral quarks.  
As in~\cite{bagan_ir}, this
cancellation then permits an expansion in the external gluon momenta {\em prior} to 
evaluating Feynman integrals (as opposed to an expansion {\em after} evaluating Feynman integrals 
used to explicitly show the IR-cancellation).

Having justified an expansion in external momenta prior to evaluating Feynman integrals, it
immediately follows that\footnote{One-particle reducible self-energy contributions to the external gluons are
ignored on both sides of~(\ref{nlo}).}
\begin{align}
  \bracket{1}{G_B^2}{2}^{(1)}
  &= \mathfig{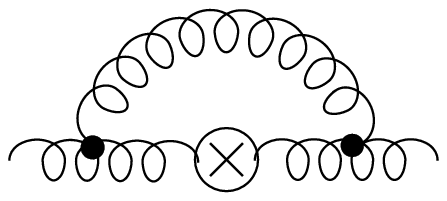}{0.6} + \mathfig{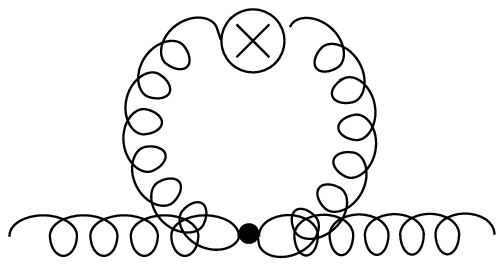}{0.6}\notag\\
  &\quad+ \mathfig{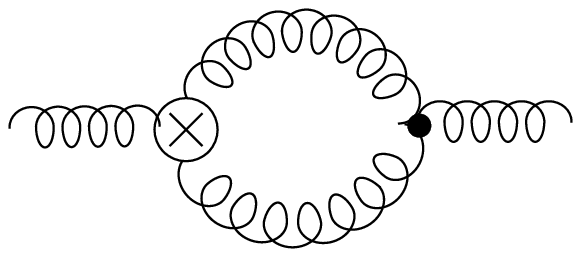}{0.6} + \mathfig{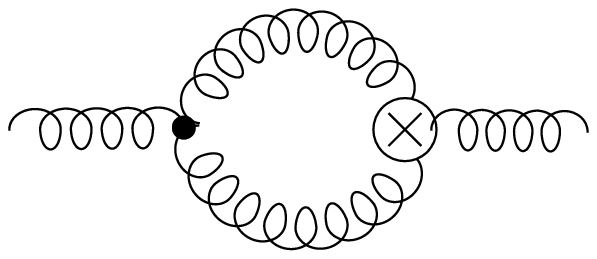}{0.6} \notag\\
  &=\Ok{4},
\end{align}
as a consequence of the resulting massless tadpole integrals, and hence~(\ref{nlo}) 
can be simplified to
\begin{equation}\label{simplified}
  \ii\int\!\dif{\hat{q}}\,\dDx\;\e^{\ii q\cdot x}
  \bracket{1}{\tprod G_B^2(x)G_B^2(0)}{2}^{(1)}
  = 4\delta^{c_1 c_2}\mc{C}^{(1)}(q^2)\polarization +\Ok{4}.
\end{equation}
Calculation of the left-hand side of~(\ref{simplified})
constitutes a sizeable project.  Fortunately, the vast majority of the required
work has already been completed in~\cite{bagan_ir}.
Therein, all contributions within the framework of purely gluonic QCD
were considered.
Therefore, to extend the result to include chiral quarks, only
those additional diagrams which admit internal quark loops need to be summed.
As previously noted,
at next-to-leading order, there are, in fact, only two such diagrams.
Thus,
\begin{align}
  &\ii\int\!\dif{\hat{q}}\,\dDx\;\e^{\ii q\cdot x}
   \bracket{1}{\tprod G_B^2(x)G_B^2(0)}{2}^{(1)}\notag\\
    &\quad=4\delta^{c_1 c_2}\frac{\alpha_{_B}}{\pi}\left\{-11\left[ \frac{1}{\hat{\varepsilon}}
      +\log\left(\frac{-q^2}{\nu^2}\right)\right]
      +\frac{131}{6}\right\}\polarization +\Ok{4}\notag\\
    &\qquad+\ii\int\dif{\hat{q}}\left\{\mathfig{Glue7.eps}{0.5}
           +{\rm crossed~diagram}\right\}
\label{fromBS}\\
    &\quad=4\delta^{c_1 c_2}\frac{\alpha_{_B}}{\pi}\left\{\left(\frac{2}{3}\nf-11\right)
       \left[ \frac{1}{\hat{\varepsilon}}
      +\log\left(\frac{-q^2}{\nu^2}\right)\right]
      +\left(\frac{131}{6}-\frac{13}{9}\nf\right)\right\}\\
    &\qquad\times\polarization+\Ok{4}\label{plusquarks}
\end{align}
where $D=4+2\varepsilon$,  $\frac{1}{\hat{\varepsilon}}=\frac{1}{\varepsilon}-\gamma_{_E} + \ln(4\pi)$
and where the first term on the right-hand side of~(\ref{fromBS}) is the purely
gluonic contribution calculated in~\cite{bagan_ir}.  Together,
Eqs.~(\ref{simplified}) and~(\ref{plusquarks}) imply that
\begin{equation}\label{c1}
  \mc{C}^{(1)}(q^2)=\frac{\alpha_{_B}}{\pi}\left\{\left(\frac{2}{3}\nf-11\right)
       \left[ \frac{1}{\hat{\varepsilon}}
      +\log\left(\frac{-q^2}{\nu^2}\right)\right]
      +\left(\frac{131}{6}-\frac{13}{9}\nf\right)\right\}.
\end{equation}
Lastly, recalling~(\ref{current}) and using~(\ref{c0}) and~(\ref{c1}), we find
\begin{align}
  &\ii\int\!\dDx\;\e^{\ii q\cdot x}\bracket{0}{\tprod J_B(x)J_B(0)}{0}\notag\\
  &\quad=\alpha_{_B}\left[\mc{C}^{(0)}(q^2)+\mc{C}^{(1)}(q^2)\right]
   \langle\alpha_{_B} G^2_B\rangle+\cdots\notag\\
  &\quad=4\alpha_{_B}\left\{ \frac{4}{D}+\frac{\alpha_{_B}}{\pi}\left[\left(\frac{\nf}{6}-\frac{11}{4}\right)
    \left(\frac{1}{\hat{\varepsilon}}+\log\left(\frac{-q^2}{\nu^2}\right)\right)
    +\left(\frac{131}{24}-\frac{13}{36}\nf\right)\right]\right\}
    \langle\alpha_{_B} G^2_B\rangle
    +\cdots
    \label{bareOPE}
\end{align}
where the dots represent contributions to the scalar gluonic correlator
arising from operators of dimension other than four.

Eq.~(\ref{bareOPE}) is expressed in terms of bare quantities and so must be
renormalized.
Renormalization of composite operators is, of course, complicated by
operator mixing~\cite{op_mix}.    Briefly,
renormalized versions (no subscript) of $\alpha$ and $G^2$  are defined as
\begin{gather}
  \alpha=Z^{-1}_{\alpha}\alpha_{_B}, \label{renorm_alpha}\\
  G^2=Z_{G^2}G^2_B +\cdots \label{renorm_G2}
\end{gather}
where the dots in~(\ref{renorm_G2}) represent contributions from equation
of motion and BRS invariant operators.  Vacuum expectation values of
these omitted operators vanish and so they do not contribute to the
calculation at hand; only $Z_{G^2}$ is actually required.
In the $\overline{\text{MS}}$ scheme the renormalization constant is \cite{op_mix}
\begin{equation}\label{Zalpha}
  Z_{G^2}=Z_{\alpha}=1+\alpi\left(\frac{11}{4}-\frac{\nf}{6}\right)
    \frac{1}{\hat{\varepsilon}};
\end{equation}
thus, $\alpha G^2$ is renormalization-group invariant at next-to-leading order.
An  analysis of the renormalization of
$G^2$ using the NZI method is presented in~\cite{bagan_sr}, where
$Z_{G^2}$ is easily distinguished
from similarly defined renormalization constants corresponding to
the omitted operators; hence $Z_{G^2}$ can be determined
by consideration of only a single amplitude: $\bracket{1}{G^2_B}{2}$.
Noting that
\begin{equation}\label{dimension}
  \frac{4}{D}=1-\frac{\varepsilon}{2},
\end{equation}
substitution of~(\ref{renorm_alpha}) and~(\ref{renorm_G2})
into~(\ref{bareOPE}) yields the following result in the $\overline{\text{MS}}$ scheme:
\begin{equation}\label{final_result}\begin{aligned}
  &\ii\int\!\dDx\;\e^{\ii q\cdot x}\bracket{0}{\tprod J(x)J(0)}{0}\\
  &\quad=\text{perturbation theory}\\
  &\quad+4\alpha\left\{ 1+\alpi\left[\left(\frac{\nf}{6}-\frac{11}{4}\right)
    \log\left(\frac{-q^2}{\nu^2}\right)
    +\left(\frac{49}{12}-\frac{5}{18}\nf\right)\right]\right\}
    \langle\alpha G^2\rangle\\
  &\quad+\text{higher dimension condensate contributions}.
\end{aligned}\end{equation}
Finally, recalling~(\ref{d4cond}) allows us to identify the coefficients
\begin{gather}
  a_0 = 4\alpha,\label{finalresultsa0}\\
  b_0 = 4\alpha\left(\frac{49}{12}-\frac{5}{18}\nf \right),\\
  b_1 = 4\alpha\left(\frac{\nf}{6}-\frac{11}{4} \right)\label{finalresultsb1}.
\end{gather}
appearing in Eq.\ (\ref{d4cond}).  Modification of these one-loop results for a change of
operator basis to $\beta(\alpha)G^2$ from $\alpha G^2$ (in the operator $J$ and/or in the OPE) can be achieved by algebraic
rearrangements.


\smallskip
\noindent
{\bf Acknowledgments:}
This research was funded through the
Office of Research Services at UCFV (DH) and the Natural Sciences \&
Engineering Council of Canada (TGS).
All Feynman diagrams were drawn using JaxoDraw 1.2-0~\cite{jaxodraw}.

\end{document}